\documentclass[twocolumn, tighten, twocolappendix, trackchanges]{aastex701}
\usepackage[utf8]{inputenc}
\usepackage[T1]{fontenc}
\usepackage{hyperref} 
\hypersetup{colorlinks=true}

\usepackage{lmodern,amsmath,amssymb,amsfonts,graphicx,float}
\usepackage{bm,siunitx,caption,subcaption,placeins}
\usepackage[dvipsnames]{xcolor}
\usepackage{CJK}

\begin{document}
\begin{CJK*}{UTF8}{gkai}

\title{Recovering Subtle Cosmological Information with the Zel'dovich-inspired transform}

\author[orcid=0000-0003-4064-417X,gname=Yun, sname=Wang]{Yun Wang (王云)}
\affiliation{Key Laboratory of Material Simulation Methods \& Software of Ministry of Education, \\ College of Physics, Jilin University, Changchun 130012, China}
\email{yunw@jlu.edu.cn}  

\author[orcid=0000-0001-5277-4882,gname=Hao-Ran, sname=Yu]{Hao-Ran Yu (于浩然)} 
\affiliation{Department of Astronomy, Xiamen University, Xiamen, Fujian 361005, China}
\email{haoran@xmu.edu.cn}

\author[orcid=0000-0002-9359-7170, gname=Yu, sname=Yu]{Yu Yu (余瑜)}
\affiliation{Department of Astronomy, \& State Key Laboratory of Dark Matter Physics,\\
            School of Physics and Astronomy, Shanghai Jiao Tong University, Shanghai 200240, China}
\affiliation{Key Laboratory for Particle Astrophysics and Cosmology (MOE),\\
\& Shanghai Key Laboratory for Particle Physics and Cosmology, Shanghai 200240, China}
\email{yuyu22@sjtu.edu.cn}

\author[orcid=0000-0001-7767-6154, gname=Ping, sname=He]{Ping He (何平)}
\affiliation{Center for Theoretical Physics, College of Physics, Jilin University, Changchun 130012, China}
\affiliation{Center for High Energy Physics, Peking University, Beijing 100871, China}
\email{hep@jlu.edu.cn}

\correspondingauthor{Yun Wang}
\email{yunw@jlu.edu.cn}

\begin{abstract}

Extracting cosmological information from nonlinear and non-Gaussian large-scale structure remains a major challenge. We introduce the Zel'dovich-inspired (ZI) transform, a simple one-parameter local nonlinear transform in which $\eta$ controls the weighting of higher-order information in the transformed density field. For $\eta\geq3$, the transform can substantially suppress gravitational non-Gaussianity. Using real-space matter density fields from the \textsc{Quijote} suite and Fisher analysis, we find that the joint data vector of three transformed-field power spectra with $\{\eta=\infty, 6, 3\}$, denoted $P_\mathrm{ZI}$, reduces marginalized errors for all parameters considered. The improvements range from several-fold for some $\Lambda$CDM parameters to factors of 290 for local primordial non-Gaussianity and 107 for the summed neutrino mass. Tests on independent fiducial simulations yield nearly unbiased parameter estimates. \textbf{Compared with other state-of-the-art summary statistics, $P_\mathrm{ZI}$ provides tighter cosmological constraints.}

\end{abstract}

\keywords{\uat{Cosmology}{343} --- \uat{Large-scale structure of the universe}{902} --- \uat{Non-Gaussianity}{1116} --- \uat{Fisher's Information}{1922} --- \uat{N-body simulations}{1083}}


\section{Introduction}\label{sec:intro}

The large-scale structure (LSS) of the Universe contains a vast amount of cosmological information. Stage-IV surveys such as DESI \citep{DESI2016}, and Euclid \citep{Amendola2018} trace the LSS over vast cosmic volumes with unprecedented precision, providing new opportunities to unravel cosmic mysteries, e.g., the neutrino mass hierarchy, the initial conditions of the Universe, and the nature of dark energy and dark matter. However, to fully release the potential of these observations, one must confront the high degree of non-Gaussianity of the late-time matter distribution induced by nonlinear gravitational evolution.

The late-time non-Gaussianity implies that cosmological information originally encoded in the power spectrum leaks into higher-order $n$-point correlations. Capturing this leaked information has become one of the central issues in current cosmological research. Simulation-based methodologies are central to addressing this challenge, with accurate and scalable cosmological simulations providing an ideal testbed for developing new tools to analyze nonlinear structure \citep[e.g.][]{Villaescusa2020,Garrison2021,Yu2026}. A wide range of tools has been developed, including the reconstruction of the initial conditions \citep[e.g.][]{Floss2024,Wang_yuting2024}, local nonlinear transformation of the late-time density field \citep[e.g.][]{Coles1991,Neyrinck2009,Joachimi2011,Simpson2011}, field-level analysis \citep[e.g.][]{Shao2023,simbig2024}, and many novel summary statistics \citep[e.g.][]{Massara2021,Valogiannis2022,Bonnaire2022,Paillas2023,Yip2024,Sunseri2025,Wang2025,Repp2025}. Among them, we focus on the power spectrum of the locally nonlinear-transformed density field because it is physically interpretable and computationally efficient, properties that are particularly valuable in the era of data-intensive cosmology. Previous studies have found that the power spectrum of the transformed field, e.g., the log power spectrum, provides moderate improvements in constraining standard cosmological parameters \citep{Neyrinck2011a,Seo2012,Wolk2015,Simpson2016}. However, its potential for constraining more subtle cosmological signatures, including the primordial non-Gaussianity (PNG) amplitudes and the sum of neutrino masses, has been less explored.

In this paper, we introduce a parametrized local nonlinear transform motivated by the Zel'dovich approximation \citep{Zeldovich1970,Shandarin1989} under special collapses. As a proof-of-concept, we apply this transform to matter density fields in real space from $N$-body simulations to remove gravitationally induced non-Gaussianity, and forecast the constraining power of the transformed-field power spectra on PNG amplitudes, summed neutrino mass, and standard $\Lambda$CDM parameters.

\section{Theory} \label{sec:theory}

The Zel'dovich approximation describes the trajectories of matter elements as linear displacements driven by the gravitational potential, providing an intuitive dynamical picture of the transition from linear to nonlinear stages. It relates the comoving Eulerian position $\boldsymbol{x}$ of a matter element at time $t$ to its unperturbed Lagrangian position $\boldsymbol{q}$ through $\boldsymbol{x}(\boldsymbol{q},t) {=} \boldsymbol{q}{-}D(t)\nabla\Phi(\boldsymbol{q},t_i)/[4\pi G\bar\rho(t_i) a^2(t_i)]$, where $D(t)$ is the linear growth factor, and $\Phi(\boldsymbol{q},t_i)$ denotes the initial gravitational potential at $t_i{\rightarrow} 0$.
By applying mass conservation for each element
from $t_i$ to $t$, and taking into account the cosmic expansion $\bar\rho\propto a^{-3}$, we get 
\begin{equation}
\label{eq:delta_za}
    1 + \delta(\boldsymbol{q},t)=\rho(\boldsymbol{x}(\boldsymbol{q}),t)/\bar\rho(t)
    = \prod_{i=1}^3[1-D(t)\alpha_i(\boldsymbol{q})]^{-1},
\end{equation}
where $\alpha_i$ are the eigenvalues of the deformation tensor $\partial^2(\Phi/4\pi G\bar\rho a^2)/\partial q_j\partial q_k$. In the linear regime, the density field can be linearized by expanding the above expression to first order
\begin{equation}
\label{eq:delta_lin}
    \delta_\mathrm{lin}(\boldsymbol{q},t) = D(t)[\alpha_1(\boldsymbol{q})+\alpha_2(\boldsymbol{q})+\alpha_3(\boldsymbol{q})].
\end{equation}
Considering the planar ($\alpha_1=\alpha, \alpha_2=\alpha_3=0$), filamentary ($\alpha_1=\alpha_2=\alpha,\alpha_3=0$), and spherical ($\alpha_1=\alpha_2=\alpha_3=\alpha$) collapses, then substituting Eq.~\eqref{eq:delta_lin} into Eq.~\eqref{eq:delta_za} yields the analytic mapping from the linear density $\delta_\mathrm{lin}$ to the nonlinear density $\delta$ as below
\begin{equation}
\label{eq:spherical_ZA}
    1+\delta=[1-\delta_\mathrm{lin}/\eta]^{-\eta}\,\, \text{with}\,\, \eta=1,2,3,
\end{equation}
which remains exact before shell-crossing occurs. This illustrates that, at the level of Zel’dovich approximation, nonlinear gravitational evolution admits an analytic inverse mapping prior to shell-crossing. Extending this idea to the fully evolved cosmic matter density field traced by galaxy surveys or outputted by numerical simulations, we can introduce a kind of nonlinear transform,
\begin{equation}
    \label{eq:Zel-kappa}
    \delta_{\mathrm{ZI}\text{-}\eta}\equiv \eta\big[1-(1+\delta)^{-1/\eta} \big],
\end{equation}
hereafter referred to as the \textit{Zel'dovich-inspired (ZI) transform}. Note that in the real universe $\eta$ need not be restricted to $\{1,2,3\}$ and is therefore treated as a free parameter. It is clear that as $\eta$ approaches infinity, Eq.~\eqref{eq:Zel-kappa} becomes a log transform, i.e., $\log(1+\delta)$.

\begin{figure}
\includegraphics[width=\columnwidth]{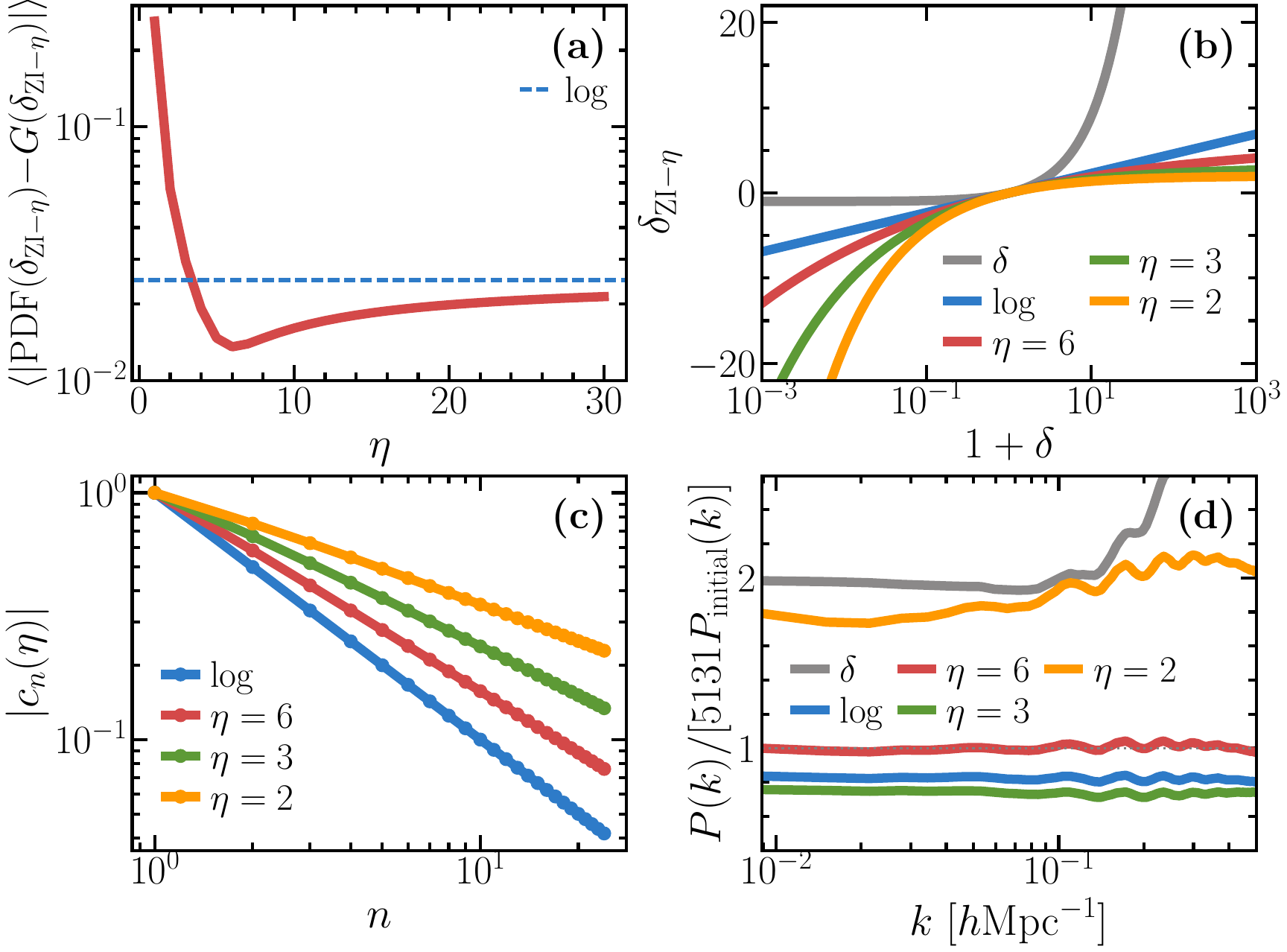}
\centering
\caption{\label{fig:kappa_pdf_ps} Gaussianization under ZI transform with different values of $\eta$. \textbf{(a)} Mean absolute deviation between the transformed density PDF and the Gaussian distribution. The blue dashed line marks the deviation in the log limit, equal to 0.0248.
\textbf{(b)} The ZI transformed field $\delta_{\mathrm{ZI}\text{-}\eta}$ versus the original density field $\delta$. 
\textbf{(c)} The Modulus of expansion coefficients of the ZI transformed field. 
\textbf{(d)} Ratios of power spectra to the initial power spectrum, which are normalized by the factor $P_{\mathrm{ZI}\text{-}6}(k_f)/P_\mathrm{initial}(k_f)=5131$ with $k_f=0.0089\, h\, \text{Mpc}^{-1}$. PDFs and spectra are averaged over 100 fields of the fiducial \textsc{Quijote} simulations (see Section~\ref{sec:data}). }
\end{figure}

To determine the optimal value of $\eta$, we measure the mean absolute deviation between the probability distribution functions (PDFs) of $\delta_{\mathrm{ZI}\text{-}\eta}$ and the Gaussian distribution using $N$-body simulations, finding that $\eta=6$ minimizes this deviation, as shown by Fig.~\ref{fig:kappa_pdf_ps} \textbf{(a)}. With this setting, the $\mathrm{ZI}$ transform yields a field whose high-density tail is more compressed than under the log transform, whereas the low-density end is pulled lower; this behavior becomes increasingly pronounced as $\eta$ is decreased (see Fig.~\ref{fig:kappa_pdf_ps} \textbf{(b)}). 

On sufficiently large scales, where the density field $\delta$ remains small, the ZI transform can be expanded in powers of $\delta$ around 0, $\delta_{\mathrm{ZI}\text{-}\eta} {=} \sum_{n=1}^{\infty}c_n(\eta)\delta^n$. Substituting this series into the power spectrum definition $\langle f(\boldsymbol{k})g(\boldsymbol{k}')\rangle=(2\pi)^3\delta^\mathrm{D}(\boldsymbol{k}+\boldsymbol{k}')P_{f,g}(k)$ gives 
\begin{equation}
    \label{eq:power_spectrum_series}
    P_{\mathrm{ZI}\text{-}\eta}(k)=\sum_{m,n}c_m(\eta)c_n(\eta)P_{\delta^m,\delta^n}(k),
\end{equation}
in which the $n$-th coefficient is 
\begin{equation}
\label{eq:Taylor_coef}
 c_n(\eta){=}\frac{(-1)^{n{+}1}}{n!}\prod_{i{=}1}^{n{-}1}(i{+}1/\eta) \ \propto \ n^{-1+1/\eta}.
\end{equation}
Eq.~\eqref{eq:power_spectrum_series} is not intended as a perturbative prediction in the nonlinear regime, where the Taylor expansion may not converge pointwise \citep{Rubira2021}. Since the coefficient $c_n(\eta)\propto n^{-1+1/\eta}$ depends explicitly on $\eta$ (see also Fig.~\ref{fig:kappa_pdf_ps} \textbf{(c)}), varying $\eta$ changes the weight of cross-spectra $P_{\delta^m,\delta^n}(k)$, with the next-to-leading-order term $P_{\delta,\delta^2}(k)=\int B(k,q,|\boldsymbol{k}-\boldsymbol{q}|)d^3\boldsymbol{q}/(2\pi^3)$ corresponding to the skew spectrum \citep{Dai2020a,Dai2020b}.  Therefore, the parameter $\eta$, acting as a \textit{higher-order information weighting parameter}, controls how bispectrum and higher-order spectra are folded into the ZI-$\eta$ power spectrum, $P_{\mathrm{ZI}\text{-}\eta}(k)$. By comparing the ZI-$\eta$ power spectra for different values of $\eta$ with the initial linear power spectrum, it can be seen from Fig.~\ref{fig:kappa_pdf_ps} \textbf{(d)} that, for $\eta \geq 3$, the transformed spectra closely follow the shape of the linear spectrum, differing primarily in amplitude.

In summary, the ZI transform with $\eta \geq 3$ can Gaussianize and linearize the nonlinear density field. This can be understood from two perspectives. Physically, the ZI transform serves as an emulator that undoes gravitational evolution. Mathematically, it transfers higher-order information into the ZI-$\eta$ power spectrum. These suggest that these power spectra may capture more information about PNG and the neutrino mass, in addition to standard cosmological parameters. Then we test this expectation via parameter forecasts on real-space matter density fields from $N$-body simulations, deferring observational complexities to a follow-up study.

\begin{table*}[t]
\caption{\label{tab:constraints}%
Marginalized and un-marginalized $1\sigma$ parameter constraints $\sigma(\theta)$ obtained with the ordinary matter power spectrum $P_\delta$, ZI-$\eta$ power spectra $P_\mathrm{log}$, $P_{\mathrm{ZI}\text{-}6}$ and $P_{\mathrm{ZI}\text{-}3}$, and their combinations $P_\mathrm{ZI}=P_\mathrm{log}\oplus P_{\mathrm{ZI}\text{-}6}\oplus P_{\mathrm{ZI}\text{-}3}$ and $P_\mathrm{all}=P_\delta\oplus P_\mathrm{ZI}$. The corresponding fully un-marginalized (i.e., holding all other parameters fixed) constraints are shown in parentheses.  }
\footnotesize
\begin{ruledtabular}
\renewcommand{\arraystretch}{1.15}
\begin{tabular}{lcccccc}
 Paras. &
$P_\mathrm{\delta}$ &
$P_\mathrm{log}$ &
$P_{\mathrm{ZI-6}}$ &
$P_{\mathrm{ZI-3}}$ &
$P_{\mathrm{ZI}}$ &
$P_{\mathrm{all}}$
\\
\colrule
$f_\text{NL}^\text{local}$ & 3289.98 (31.48)  & 55.88 (6.20)    & 47.82 (5.51)     & 67.12 (7.68)   & 11.37 (3.95)  & 11.05 (3.89) \\
$f_\text{NL}^\text{equil}$ & 8239.73 (79.67)  & 3981.99 (22.91)   & 3951.96 (21.02)    & 3700.63 (27.85)  & 87.67 (17.80)   & 75.62 (17.59) \\
$f_\text{NL}^\text{ortho}$ & 3691.67 (188.28) & 1487.97 (12.24)   & 1628.88 (10.75)    & 1593.72 (14.79)  & 98.47 (8.66)   & 83.93 (8.48) \\
$M_\nu$                    & 1.18 (0.092)  & 0.75 (0.0022)  & 0.69 (0.0017)   & 0.82 (0.0031) & 0.011 (0.0011)  & 0.0079 (0.0010) \\
$h$                        & 0.49 (0.011)  & 0.50 (0.0023)  & 0.51 (0.0020)   & 0.52 (0.0027)  & 0.072 (0.0016)   & 0.069 (0.0016) \\
$n_s$                      & 0.52 (0.0089)  & 0.46 (0.0015)  & 0.47 (0.0013)  & 0.49 (0.0018)  & 0.032 (0.00099)   & 0.026 (0.00097) \\
$\Omega_m$                 & 0.090 (0.0042)  & 0.093 (0.00084)  & 0.098 (0.00074)  & 0.094 (0.00098)  & 0.011 (0.00062)   & 0.0096 (0.00061)  \\
$\Omega_b$                 & 0.039 (0.0024)  & 0.040 (0.00061)  & 0.039 (0.00054)  & 0.041 (0.00070)  & 0.0082 (0.00045)  & 0.0080 (0.00044) \\
$\sigma_8$                 & 0.093 (0.0021)  & 0.32 (0.00061)  & 0.36 (0.00057)  & 0.27 (0.00066)  & 0.0021 (0.00049)  & 0.0020 (0.00049)
\end{tabular}
\end{ruledtabular}
\end{table*}

\section{Fisher formalism} \label{sec:fisher}

We use the Fisher information formalism \citep{Fisher1922,Tegmark1997} to quantify and benchmark how much information content can be recovered by the power spectra after applying the ZI transform with three representative choices of $\eta$: $\eta=\infty$ for the log transform, $\eta=6$ yielding the most Gaussianized PDF, and $\eta=3$ corresponding to the inverse Zel'dovich approximation in the spherical collapse. Reasonably assuming a Gaussian likelihood function \citep{Upham2021,Paillas2023,Ouellette2025}, the Fisher matrix reads
\begin{equation}
\label{eq:fisher}
    \mathcal{F}_{ij}=\frac{\partial \boldsymbol{d}}{\partial \theta_i}\mathcal{C}^{-1}\frac{\partial \boldsymbol{d}}{\partial \theta_j}^\mathrm{T},
\end{equation}
where $\boldsymbol{{d}}$ is the data vector consisting of the ordinary power spectrum and ZI-$\eta$ power spectra, $\theta$ is the collection of cosmological parameters, and $\mathcal{C}$ is the covariance matrix independent of cosmology (see Appendix~\ref{sec:cov_deriv} for more details). In this framework, the Cram\'er–Rao theorem sets a lower bound on the marginalized $1\sigma$ error of parameter $\theta_i$, i.e. $\sigma(\theta_i)\geq \sqrt{(\mathcal{F}^{-1})_{ii}}$.

\section{Data set} \label{sec:data}

The accurate estimation of the Fisher matrix requires a large ensemble of simulations, which is fulfilled by the publicly available \textsc{Quijote} and its extension \textsc{Quijote-PNG} suites (hereafter \textsc{Quijote}) \citep{Villaescusa2020,Jung2022,Coulton2023a}. They provide tens of thousands of $N$-body simulations spanning a wide cosmological parameter space, which are organized into several subsets, e.g., the fiducial set of 15,000 random realizations with Planck cosmology (14,500 of which are used for covariance estimation), and additional sets of 500 realizations, each set with a single parameter perturbed around its fiducial value for computing derivatives. The simulations evolve $512^3$ dark matter particles (and $512^3$ neutrinos for massive-neutrino runs) in a cubic volume of $1\ (h^{-1}\mathrm{Gpc})^3$ from redshift $z=127$ to the present, using the \textsc{TreePM} code \textsc{GADGET-III} with initial conditions from second-order Lagrangian perturbation theory (and Zel'dovich approximation for massive-neutrino runs). In this work, the matter density field (dark matter for massless-neutrino runs and total matter for massive-neutrino runs) is constructed on a $512^3$ grid using the PCS scheme \citep{Sefusatti2016}, and analyzed with the public \textsc{Pylians} library \citep{Pylians}. We focus on the matter density fields at redshift $z=0$, and consider nine cosmological parameters consisting of the summed neutrino mass $M_\nu=\sum_im_i$, the PNG amplitudes $\{f^\mathrm{local}_\mathrm{NL}$, $f^\mathrm{equil}_\mathrm{NL}$, $f^\mathrm{ortho}_\mathrm{NL}\}$, and the $\Lambda$CDM parameters $\{h$, $n_s$, $\Omega_m$, $\Omega_b$, $\sigma_8\}$.

\section{Results} \label{sec:results}

Table~\ref{tab:constraints} presents the marginalized and un-marginalized constraints for each parameter obtained from the matter power spectrum $P_\delta$, the log power spectrum $P_\mathrm{log}$, the ZI-6 power spectrum $P_{\mathrm{ZI}\text{-}6}$, the ZI-3 power spectrum $P_{\mathrm{ZI}\text{-}3}$,  the combination of ZI transformed spectra $P_\mathrm{ZI}=P_\mathrm{log}\oplus P_{\mathrm{ZI}\text{-}6}\oplus P_{\mathrm{ZI}\text{-}3}$, and the combination of all spectra $P_\mathrm{all}=P_\delta\oplus P_\mathrm{ZI}$. \textbf{The wavenumber range included in our analysis is $0.0089\, h\, \text{Mpc}^{-1}\leq k\leq 0.5\, h\mathrm{Mpc}^{-1}$, with the maximum wavenumber chosen to avoid numerical artifacts.} We have verified that the constraints from $P_\mathrm{log}$, $P_{\mathrm{ZI}\text{-}6}$, $P_{\mathrm{ZI}\text{-}3}$, $P_\mathrm{ZI}$, and $P_\mathrm{all}$ are numerically converged, although $P_\delta$ alone does not converge with respect to the derivative estimates (Appendix~\ref{sec:converge}).

We first compare the constraining power of the individual ZI-$\eta$ power spectrum with that of $P_\delta$. For the fully marginalized constraints, all transformed spectra improve the constraints on $M_\nu$ and the PNG amplitudes relative to $P_\delta$. The improvement is modest for $M_\nu$, $f_\mathrm{NL}^\mathrm{equil}$, and $f_\mathrm{NL}^\mathrm{ortho}$, but is particularly significant for $f_\mathrm{NL}^\mathrm{local}$, where the errors are reduced by factors of tens. In contrast, the transformed spectra provide little to no improvement for the standard $\Lambda$CDM parameters, with the most severe degradation occurring for $\sigma_8$. The unmarginalized constraints show a different behavior: each ZI-$\eta$ spectrum outperforms $P_\delta$ for all parameters considered, with $P_{\mathrm{ZI}\text{-}6}$ generally giving the tightest constraints, followed by $P_\mathrm{log}$ and then $P_{\mathrm{ZI}\text{-}3}$. 

\begin{figure}
\includegraphics[width=\columnwidth]{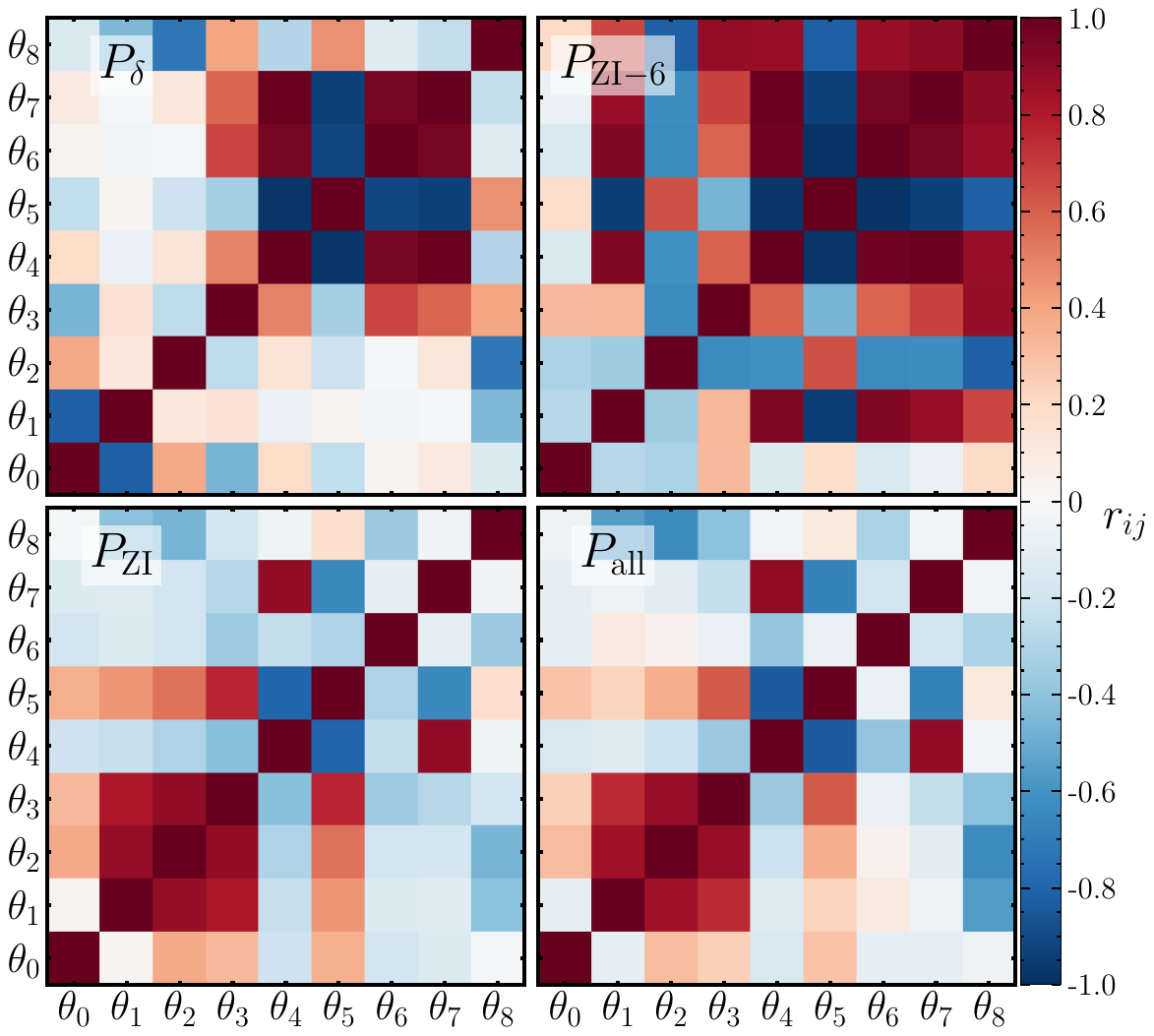}
\centering
\caption{\label{fig:parameter_correlation} Parameter degeneracy quantified by the correlation $r_{ij}=(\mathcal{F}^{-1})_{ij}/\sqrt{(\mathcal{F}^{-1})_{ii}(\mathcal{F}^{-1})_{jj}}$ between parameters $\theta_i$ and $\theta_j$ for different power spectra and combinations, where $\theta_i\in\{f_\text{NL}^\text{local},f_\text{NL}^\text{equil}, f_\text{NL}^\text{ortho}, M_\nu, h, n_s, \Omega_m, \Omega_b, \sigma_8 \}$. The panels for $P_{\mathrm{log}}$ and $P_{\mathrm{ZI}\text{-}3}$ are omitted, since their correlation patterns are visually similar to that of $P_{\mathrm{ZI}\text{-}6}$.}
\end{figure}

This difference between marginalized and un-marginalized constraints can be understood as follows. The ZI transform can eliminate nearly all off-diagonal elements of the covariance induced by nonlinear gravitational mode coupling (see Fig.~\ref{fig:spectra_correlation}). Consequently, the unmarginalized information is almost completely governed by the derivative of the normalized power spectrum, $\partial [P(k)/\sqrt{\mathcal{C}(k,k)}]/\partial\theta$ (Eq.~\ref{eq:fisher_diag}). Among the individual transformed spectra, this response is typically strongest for $P_{\mathrm{ZI}\text{-}6}$, followed by $P_\mathrm{log}$ and $P_{\mathrm{ZI}\text{-}3}$, explaining the ordering of the unmarginalized constraints (see Fig.~\ref{fig:normalized_derivatives}). Once all parameters are marginalized over, parameter degeneracies become important because different parameters can have similar derivative shapes, which are demonstrated clearly in Figs.~\ref{fig:parameter_correlation} and \ref{fig:normalized_derivatives}. Since the ZI-$\eta$ spectra retain enhanced responses to $M_\nu$ and PNG amplitudes, the marginalized gains are concentrated in them, while the constraints on $\Lambda$CDM parameters degrade, with $n_s$ being the exception. 

When the three ZI-$\eta$ power spectra are combined, the marginalized constraints are tightened for all parameters. Most notably, the constraining power of $P_\mathrm{ZI}$ relative to $P_\delta$ is boosted by factors of $\sim290$, $\sim107$, and $\sim45$ for $f_\mathrm{NL}^\mathrm{local}$, $M_\nu$, and $\sigma_8$, respectively. We also see that, in Fig.~\ref{fig:parameter_correlation}, $P_\mathrm{ZI}$ substantially reduce parameter degeneracies, e.g. the correlation between $\sigma_8$ and $h$ is only $-0.04$. Adding $P_\delta$ further to the joint analysis leads to only a minor improvement, indicating that most information gain is already captured by $P_\mathrm{ZI}$. The outstanding performance arises from the joint covariance structure and distinct parameter responses. Because the transformed spectra are highly correlated at the same $k$ but weakly correlated across different $k$-modes, the inverse covariance suppresses their nearly common fluctuations and enhances their differences. These differences are induced by the $\eta$-dependent weighting of higher-order information and lead to distinct parameter responses, as shown in Figs.~\ref{fig:normalized_derivatives} and \ref{fig:normalized_derivatives_diff_comps}. For the formal description, see Appendix~\ref{sec:cov_deriv}.

As a benchmark, we next compare the constraints from $P_\mathrm{ZI}$ with those from other advanced summary statistics based on the same simulation suite, including the bispectrum, wavelet scattering transform \citep{Valogiannis2022}, marked power spectrum \citep{Massara2021}, cosmic-web environment-dependent power spectra \citep{Bonnaire2022,Sunseri2025}, halo mass function and void size function \citep{Bayer2021}, and scale-dependent peak/valley statistics \citep{Wang2025}. We find that $P_\mathrm{ZI}$ provides highly competitive constraints, yielding the tightest errors among the statistics considered for $f_\mathrm{NL}^\mathrm{local}$, $f_\mathrm{NL}^\mathrm{equil}$, $M_\nu$, and many standard $\Lambda$CDM parameters (see Appendix~\ref{sec:comparison} for details).

\begin{figure}[t]
\includegraphics[width=\columnwidth]{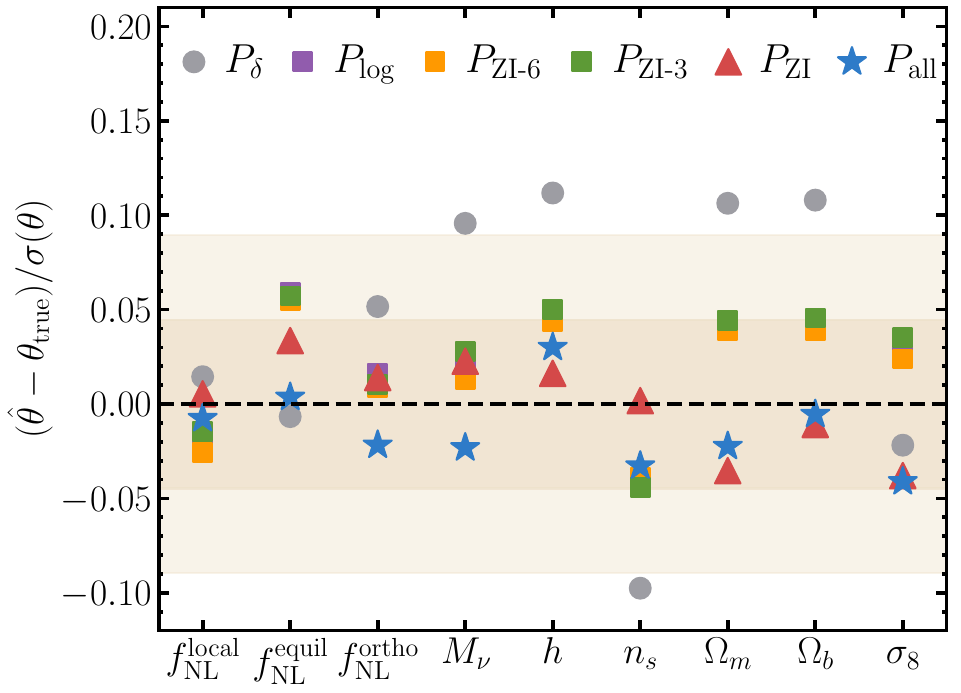}
\centering
\caption{\label{fig:normalized_bias} Normalized bias of the Fisher parameter estimators constructed from different power spectra and combinations. It is defined as $(\hat{\theta}-\theta_{\rm true})/\sigma(\theta)$, where $\hat\theta$ is the estimated parameter averaged over 500 fiducial simulations excluded from the
covariance estimation, $\theta_{\rm true}$ is the input true parameter, and $\sigma(\theta)$ is the marginalized uncertainty. \textbf{The dark and light shaded regions indicate the expected $1\sigma$ and $2\sigma$ statistical fluctuations of the mean estimate, respectively, corresponding to $\pm 1/\sqrt{500}$ and $\pm 2/\sqrt{500}$.}}
\end{figure}

To test whether the parameter estimates inferred from our power spectra and combinations are biased, we construct a parameter estimator following \citet{Alsing2018} as below

\begin{equation}
    \hat{\theta}_i=\theta_i^{\rm fid}
+\Sigma_j \left(F^{-1}\right)_{ij}
(\partial \bar{\boldsymbol{d}}/\partial \theta_j)
\mathcal{C}^{-1}
\left(\boldsymbol{d}_{\rm obs}-\bar{\boldsymbol{d}}_{\rm fid}\right),
\end{equation}
where $\boldsymbol{d}_{\rm obs}$ is the mock observed data vector, $\bar{\boldsymbol{d}}$ is the mean over 500 derivative simulations, and $\bar{\boldsymbol{d}}_{\rm fid}$ is the mean over 14,500 fiducial simulations. We apply this estimator to 500 fiducial simulations excluded from the covariance estimation. In Fig.~\ref{fig:normalized_bias}, we show the biases of the estimated parameters with respect to the true parameters of the observed data. It is observed that $P_\delta$ produces appreciable biases, with many parameters lying outside the $2\sigma$ region. In contrast, the ZI-$\eta$ spectra mostly remain within the $1\sigma$ region, and their combination $P_\mathrm{ZI}$ further reduces the biases.

\section{Conclusions} \label{sec:conc}

We derive a physically motivated local nonlinear transformation, the ZI transform (Eq.~\ref{eq:Zel-kappa}), starting from the Zel’dovich approximation under special collapse cases. This transform emulates the inverse of nonlinear gravitational evolution, while the parameter $\eta$ controls the weighting of higher-order information encoded in the transformed field. Using the real-space matter density fields of \textsc{Quijote} simulations, we see that the ZI transform with $\eta \geq3$ can restore the non-Gaussian density field toward Gaussianity and recover the linear power spectrum. With Fisher analysis, we show that the power spectra of the transformed fields with $\eta=3$, $6$, and $\infty$ (the last corresponding to the log transform) substantially improve the extraction of cosmological information from nonlinear scales. While individual ZI-$\eta$ power spectra already enhance the sensitivity to $M_\nu$ and PNG amplitudes, their combination, $P_\mathrm{ZI}$, provides much tighter constraints and breaks parameter degeneracies. Compared with several summary statistics that also go beyond the traditional power spectrum, the combination of ZI-$\eta$ power spectra achieves the best constraints for nearly all parameters. We find that the substantial improvement of $P_\mathrm{ZI}$ is not limited to precision but also extends to accuracy, as the parameter estimator constructed from it is nearly unbiased.

Our results demonstrate that power spectra of local nonlinear transformed fields offer a computationally efficient and physically interpretable path to recovering information otherwise hidden by nonlinear gravitational evolution. Given the substantial gains in both precision and estimator accuracy, ZI-$\eta$ power spectra should be further developed systematically, with theoretical and observational uncertainties gradually incorporated, and with particular attention to how primordial non-Gaussianity and massive neutrinos affect the spectra across different choices of $\eta$. Such a development has the potential to open a promising route to extracting subtle cosmological signatures from the large-scale structure.

\begin{acknowledgments}
We thank Mark Neyrinck and the anonymous referee for helpful suggestions and comments. We would also like to thank Francisco Villaescusa-Navarro, William Coulton, and the whole \textsc{Quijote} team for making their simulations publicly available. This work was supported by the Start-up Funding from Jilin University (NO. 419080525A84), and the Young Scientists Fund (C) of the National Natural Science Foundation of China (NO.~12603010).
\end{acknowledgments}

%



\appendix

\begin{figure}
\includegraphics[trim={0 0 0 25pt},clip,width=\columnwidth]{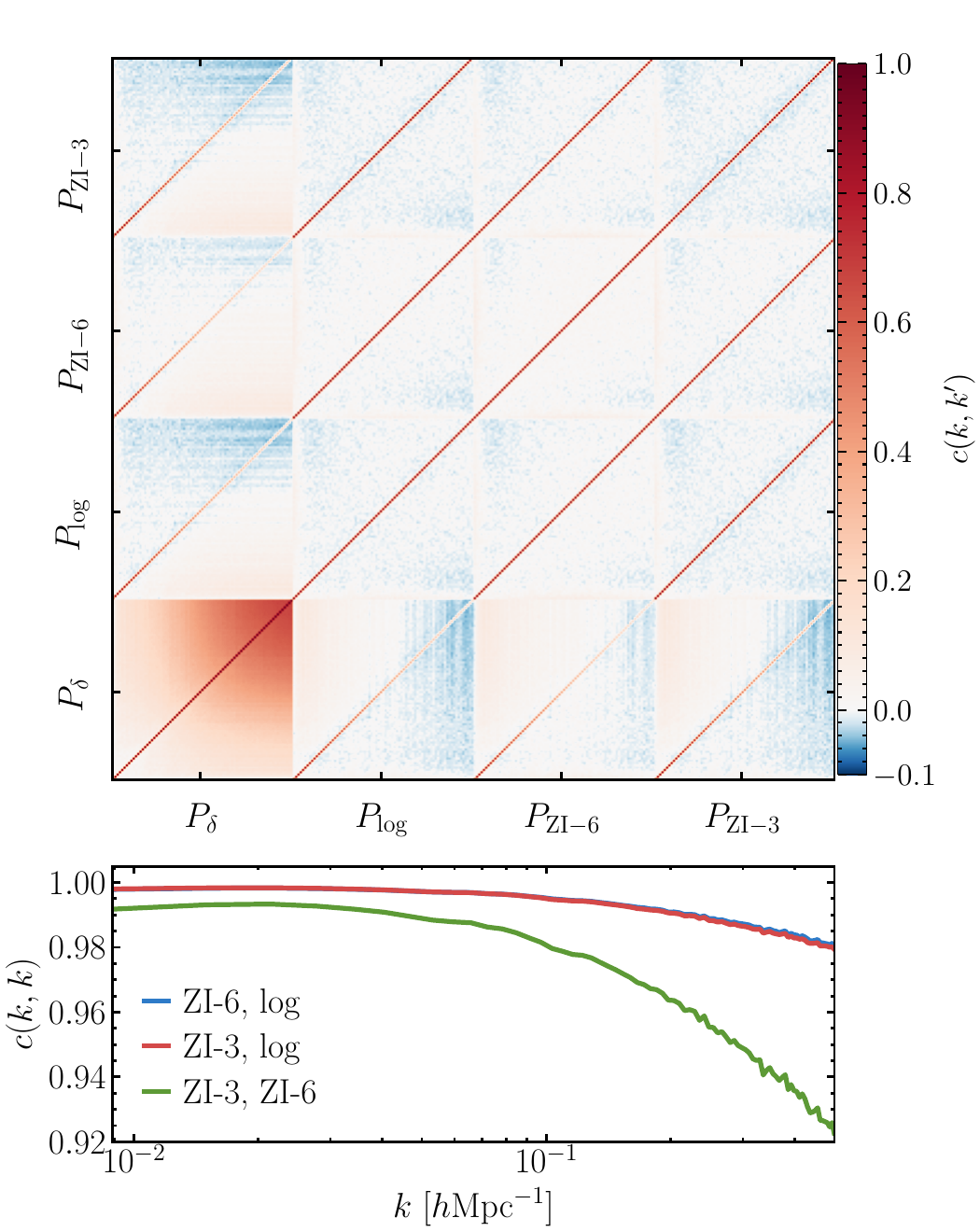}
\centering
\caption{\label{fig:spectra_correlation} Correlation coefficients $c(k,k^\prime)$ $=$ $\mathcal{C}(k,k^\prime)/$ $\sqrt{\mathcal{C}(k,k)\mathcal{C}(k^\prime,k^\prime)}$. The upper panel shows the full matrix, with each block corresponding to a pair of spectra going from $0.0089\ h\mathrm{Mpc}^{-1}$ to $0.5\ h\mathrm{Mpc}^{-1}$. The lower panel shows the same-$k$ cross-correlation coefficients between the transformed spectra, corresponding to the diagonal elements of the off-diagonal blocks in the upper panel.}
\end{figure}

\begin{figure}
\includegraphics[width=\columnwidth]{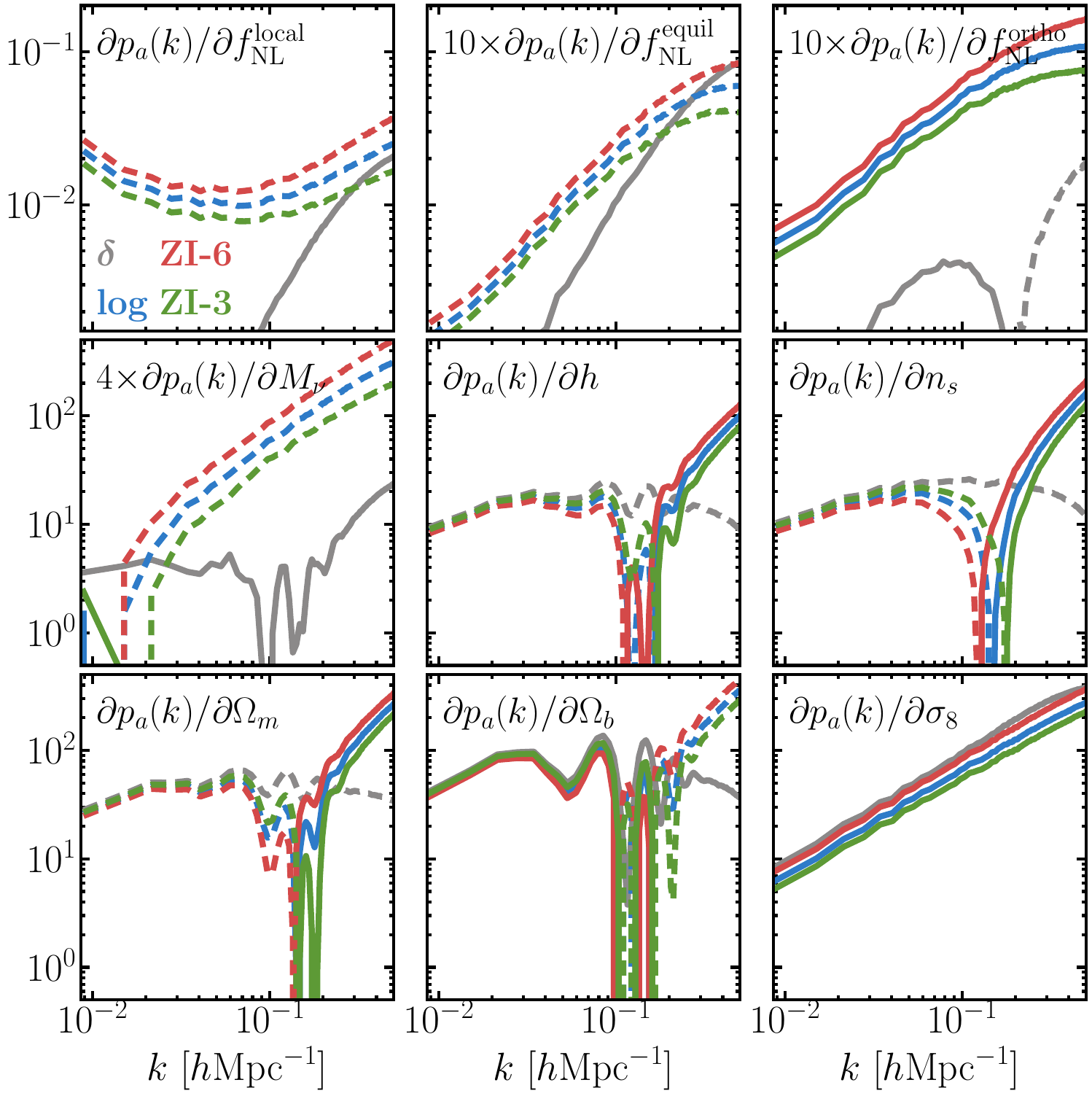}
\centering
\caption{\label{fig:normalized_derivatives} Numerical partial derivatives of $p_a(k)$ $=$ $P_a(k)/\sigma_a(k)$ $=$ $P_a(k)/\sqrt{\mathcal{C}(k,k)}$ with respect to different cosmological parameters, where ``$a$'' denotes ``$\delta$'', ``log'', ``ZI-6'', or ``ZI-3'', as labeled. The absolute values of the 
derivatives are shown; solid and dashed lines indicate positive and negative 
signs, respectively.}
\end{figure}

\section{Covariance and derivatives}
\label{sec:cov_deriv}

As seen in Eq.~\eqref{eq:fisher}, the Fisher matrix comprises two parts: the inverse of the covariance matrix and the partial derivatives of the data with respect to parameters. The covariance and its inverse are given by
\begin{equation}
\label{eq:data_cov}
\mathcal{C} {=} \langle(\boldsymbol{d}{-}\langle\boldsymbol{d}\rangle)^\mathrm{T}(\boldsymbol{d}{-}\langle\boldsymbol{d}\rangle) \rangle\ \text{and}\ 
\mathcal{C}^{-1}{\leftarrow} \frac{N_\mathrm{fid}{-}N_\mathrm{d}{-}2}{N_\mathrm{fid}{-}1}\mathcal{C}^{{-}1},\nonumber
\end{equation}
where the Hartlap factor $(N_\mathrm{fid}-N_\mathrm{d}-2)/(N_\mathrm{fid}-1)$ is multiplied to correct the finite-simulation bias \citep{Hartlap2007}. Here $N_\mathrm{fid}=14,500$ denotes the number of fiducial simulations used for covariance estimation, and $N_\mathrm{d}$ is the dimension of the data vector with $N_\mathrm{d}=79$ for each individual power spectrum.

For PNG amplitudes and $\Lambda$CDM parameters, we compute the parameter derivatives by the central difference formula:
\begin{equation}
    \frac{\partial \boldsymbol{d}}{\partial \theta_i} = \frac{\boldsymbol{d}(\theta_i+\Delta\theta_i)-\boldsymbol{d}(\theta_i-\Delta\theta_i)}{2\Delta\theta_i}.\nonumber
\end{equation}
For the summed neutrino mass $M_\nu$ that must be positive, we compute the derivative by the four-point forward approximation:
\begin{equation}
    \frac{\partial \boldsymbol{d}}{\partial M_\nu} {=}\frac{\boldsymbol{d}(4\mathrm{d}M_\nu){-}12\boldsymbol{d}(2\mathrm{d}M_\nu){+}32\boldsymbol{d}(\mathrm{d}M_\nu){-}21\boldsymbol{d}(M_\nu{=}0)}{12\mathrm{d}M_\nu}.\nonumber
\end{equation}

\begin{figure}[t]
\includegraphics[width=\columnwidth]{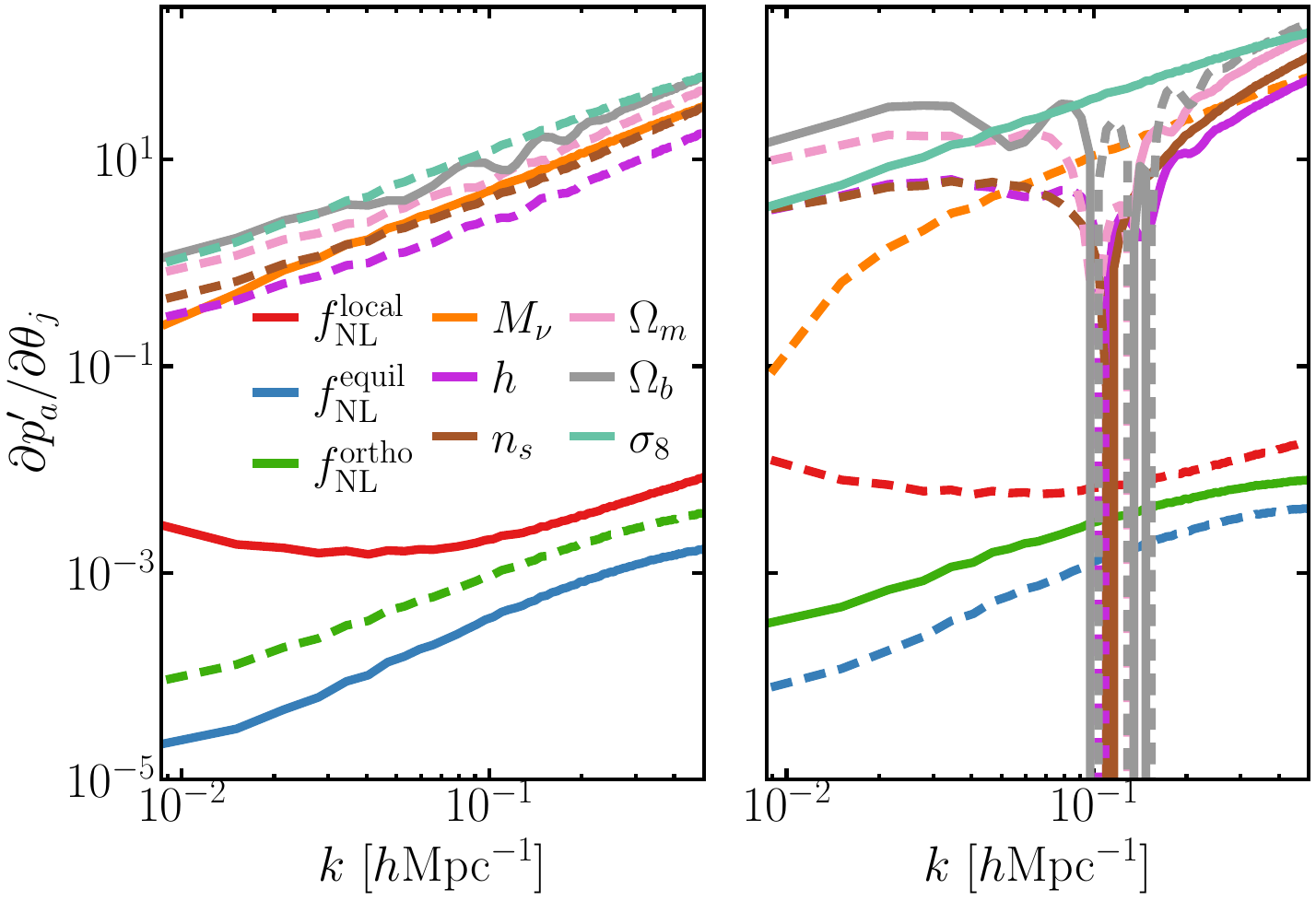}
\centering
\caption{\label{fig:normalized_derivatives_diff_comps} Derivatives of the two differential components, $p^\prime_1
$ $=$ $(p_\mathrm{log}{-}p_{\mathrm{ZI}\text{-}6})/\sqrt{2}$ (left) and $p^\prime_2$ $=$ $(p_\mathrm{log}{+}p_{\mathrm{ZI}\text{-}6}{-}2p_{\mathrm{ZI}\text{-}3})/\sqrt{6}$ (right), with respect to the cosmological parameters considered. Solid (dashed) lines indicate that the derivative is positive (negative).}
\end{figure}

Fig.~\ref{fig:spectra_correlation} shows the normalized covariance matrix of the four power spectra we considered. For the ordinary matter power spectrum $P_\delta$, Fourier modes are strongly correlated with each other on small scales due to the nonlinear gravitational evolution \citep{Blot2015}. After performing local nonlinear transformations, mode correlations are substantially suppressed, rendering the covariances of $P_\mathrm{log}$, $P_{\mathrm{ZI}\text{-}6}$, and $P_{\mathrm{ZI}\text{-}3}$ nearly diagonal. For each ZI-$\eta$ power spectrum, the Fisher matrix can therefore be simplified as
\begin{equation}
\label{eq:fisher_diag}
    \mathcal{F}_{a,ij}\approx\sum_{n=0}^{78}\frac{\partial p_a(k_n)}{\partial \theta_i}\frac{\partial p_a(k_n)}{\partial \theta_j},
\end{equation}
where $p_a(k_n){=}P_a(k_n)/\sigma_a(k_n)$ is the power spectrum normalized by its standard deviation, and the subscript ``$a$'' denotes ``log'', ``ZI-6'', or ``ZI-3''. The measurements of $\partial p_a(k)/\partial\theta_i$ are shown in Fig.~\ref{fig:normalized_derivatives}, from which we see that $|\partial p_{\mathrm{ZI}\text{-}6}(k)/\partial\theta_i|$ $>$ $|\partial p_\mathrm{log}(k)/\partial\theta_i|$ $>$ $|\partial p_{\mathrm{ZI}\text{-}3}(k)/\partial\theta_i|$ holds over most scales for all parameters. These differences originate from the $\eta$-dependent nonlinear weighting of higher-order information in the ZI transform, and can be further amplified through linear superpositions of the transformed spectra, as shown in Fig.~\ref{fig:normalized_derivatives_diff_comps}. In addition, for each transformed spectrum, the derivatives with respect to the PNG amplitudes and the summed neutrino mass are notably larger than those of $P_\delta$. For $h$, $n_s$, $\Omega_m$, and $\Omega_b$, the transformed spectra exceed $P_\delta$ only on scales of $k\gtrsim0.25\ h\mathrm{Mpc}^{-1}$, whereas the response to $\sigma_8$ is weaker than that of $P_\delta$ over the full scale range. Another feature is that, except for $f_\mathrm{NL}^\mathrm{local}$, the derivatives often show clear alignment or anti-alignment, suggesting appreciable positive or negative parameter degeneracies, as illustrated in Fig.~\ref{fig:parameter_correlation}.

When combining the three ZI-$\eta$ power spectra, the full covariance matrix includes the cross-covariances among them. Fig.~\ref{fig:spectra_correlation} shows that different transformed spectra are virtually uncorrelated across different $k$ bins, but highly correlated at the same $k$. Quantitatively, the same-$k$ correlations with $P_\mathrm{log}$ decrease from $\simeq0.998$ on the largest scales to $\simeq0.98$ at $k_\mathrm{max}$, whereas the $P_{\mathrm{ZI}\text{-}3} - P_{\mathrm{ZI}\text{-}6}$ correlation falls from $\simeq0.99$ to $\simeq0.92$. Consequently, the joint covariance matrix of the three transformed spectra can be approximately reorganized as a block diagonal matrix, i.e.
\begin{equation}
    \mathcal{C} = \mathrm{diag}(C_0,\ C_1,\ C_2,\ ...,\ C_n,\ ..., C_{78}),
\end{equation}
with $C_n=D_nR_nD_n$ being
\begin{equation}
{\begin{pmatrix}
\sigma_\mathrm{log} &  & \\
 & \sigma_{\mathrm{ZI}\text{-}6} &  \\
 &  & \sigma_{\mathrm{ZI}\text{-}3}
\end{pmatrix}
\begin{pmatrix}
1 & c & c^\prime\\
c & 1 & c^{\prime\prime} \\
c^\prime & c^{\prime\prime} & 1
\end{pmatrix}
\begin{pmatrix}
\sigma_\mathrm{log} &  & \\
 & \sigma_{\mathrm{ZI}\text{-}6} &  \\
 &  & \sigma_{\mathrm{ZI}\text{-}3}
\end{pmatrix}},\nonumber
\end{equation}
where all elements of $C_n$ are computed at wavenumber of $k=k_n$. Here, $c(k_n)$, $c^\prime(k_n)$, and $c^{\prime\prime}(k_n)$ are the same-$k$ cross-correlation coefficients of $P_\mathrm{log}-P_{\mathrm{ZI}\text{-}6}$, $P_\mathrm{log}-P_{\mathrm{ZI}\text{-}3}$, and $P_{\mathrm{ZI}\text{-}6}-P_{\mathrm{ZI}\text{-}3}$, respectively. Under this approximation, the Fisher matrix can be written as
\begin{equation}
    \mathcal{F}_{ij}\approx\sum_{n=0}^{78}\frac{\partial \boldsymbol{p}(k_n)}{\partial\theta_i}R_n^{-1}\frac{\partial \boldsymbol{p}(k_n)^\mathrm{T}}{\partial\theta_j},
\end{equation}
in which $\boldsymbol{p}(k_n)=\left(\ p_\mathrm{log}(k_n),\ p_{\mathrm{ZI}\text{-}6}(k_n),\ p_{\mathrm{ZI}\text{-}3}(k_n)\ \right)$ can be viewed as a vector with three components. If we further approximate $c(k_n)=c^\prime(k_n)=c^{\prime\prime}(k_n)$, the Fisher matrix can be further simplified as
\begin{equation}
    \mathcal{F}_{ij}\approx\sum_{a=0}^{2}\sum_{n=0}^{78}\frac{1}{\lambda_{n,a}}\frac{\partial p_a^\prime(k_n)}{\partial\theta_i}\frac{\partial p_a^\prime(k_n)}{\partial\theta_j},
\end{equation}
where the new vector $\boldsymbol{p}^\prime$ comprises three components:
\begin{align}
    p^\prime_0 &= (p_\mathrm{log}{+}p_{\mathrm{ZI}\text{-}6}{+}p_{\mathrm{ZI}\text{-}3})/\sqrt{3},\nonumber\\
    p^\prime_1 &= (p_\mathrm{log}{-}p_{\mathrm{ZI}\text{-}6})/\sqrt{2},\nonumber \\
    p^\prime_2 &= (p_\mathrm{log}{+}p_{\mathrm{ZI}\text{-}6}{-}2p_{\mathrm{ZI}\text{-}3})/\sqrt{6},\nonumber
\end{align}
and $\{\lambda_{n,0}=1+2c(k_n)$, $\lambda_{n,1}=\lambda_{n,2}=1-c(k_n)\}$ are eigenvalues of $R_n$.

\begin{figure}[t]
\includegraphics[width=\columnwidth]{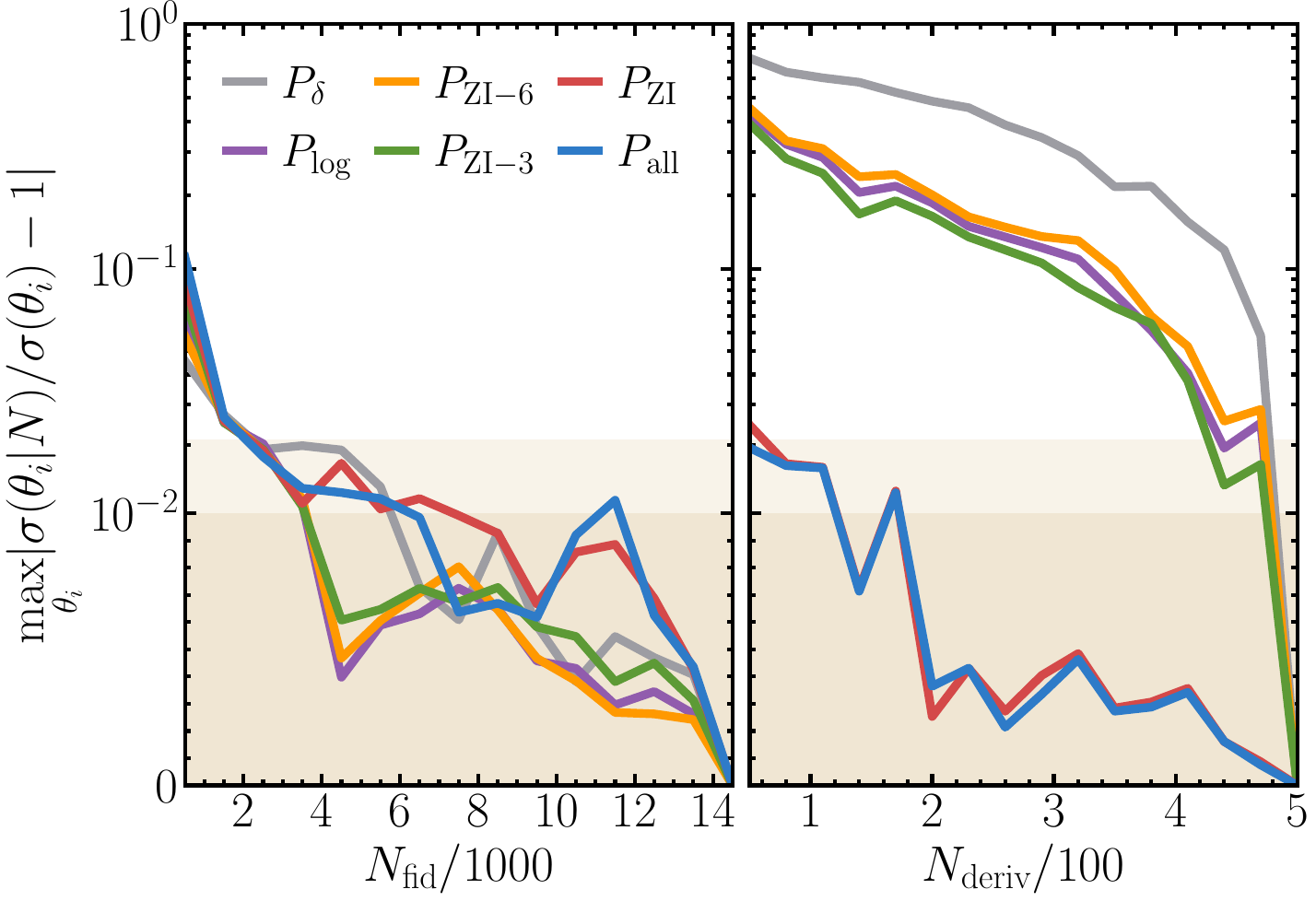}
\centering
\caption{\label{fig:convergence_test} Convergence analysis of the marginalized parameter constraints for power spectra and their combinations. The left and right panels show the convergence metric versus $N_{\rm fid}$ for covariance estimation and $N_{\rm deriv}$ for numerical derivatives, respectively. The dark (light) shaded regions indicate the $1\%$ ($2\%$) agreement level.}
\end{figure}

Since $c(k_n)$ is close to unity over the scales considered, the derivatives of the two difference components $p^\prime_1$ and $p^\prime_2$ are strongly amplified by $1/(1-c(k_n))\sim \mathcal{O}(10)-\mathcal{O}(10^2)$, while the common component $p^\prime_0$ is downweighted by $1/(1+2c(k_n))\sim1/3$. The Fisher information is therefore expected to be dominated by $p^\prime_1$ and $p^\prime_2$. As shown in Fig.~\ref{fig:normalized_derivatives_diff_comps}, the derivatives $\partial p_1^\prime/\partial\theta_i$ and $\partial p_2^\prime/\partial\theta_i$ show completely different scale dependences for a given parameter. They therefore provide complementary Fisher information. The combination of them, equivalent to the joint analysis of the original three ZI-$\eta$ power spectra, can substantially reduce parameter degeneracies and constraint errors.

\section{Convergence of Fisher constraints}
\label{sec:converge}

Following \citep{Bonnaire2022}, we use a convergence metric to assess the stability of the Fisher constraint errors obtained from different power spectra and their combinations. The metric is defined as
\begin{equation}
\max_{\theta_i}\left|\frac{\sigma(\theta_i|N)}{\sigma(\theta_i)}-1\right|,
\end{equation}
namely the maximum fractional deviation of the marginalized constraints over all cosmological parameters relative to the full-sample result that presented in the main text. Here $N$ is either $N_\mathrm{fid}$, the number of simulations used to compute
the covariances, or $N_\mathrm{deriv}$, the number of simulations used to estimate derivatives. The results of this convergence test are shown in Fig.~\ref{fig:convergence_test}.

We see that the convergence with respect to $N_{\rm fid}$ is rapid. All statistics are below the $2\%$ level when $\sim4,000$ fiducial simulations are used, and reach the percent level when $N_{\rm fid}\gtrsim6,000$. The convergence with respect to $N_\mathrm{deriv}$ is statistic-dependent. The combined spectra $P_\mathrm{ZI}$ and $P_\mathrm{all}$  are already stable at the percent level with $N_\mathrm{deriv}\sim200$. The individual transformed spectra converge more slowly, reaching the $5\%$ level at $N_\mathrm{deriv}\sim400$ and the $2\%$ level at $N_\mathrm{deriv}\sim450$. By contrast, the ordinary power spectrum $P_\delta$ does not converge over the tested range of $N_\mathrm{deriv}$, consistent with previous analyses \citep[e.g.][]{Coulton2023a,Floss2024}.

\section{Comparison with other summary statistics}
\label{sec:comparison}

Given the strong constraining power of the combination of ZI-$\eta$ power spectra $P_\mathrm{ZI}$, it is natural to compare its performance with that of the bispectrum and other recently developed summary statistics. The marginalized constraints obtained from different statistics are summarized in Table~\ref{tab:constraints_png} and \ref{tab:constraints_mnu}, where the smallest constraint error for each parameter is highlighted in bold. In the PNG+$\Lambda$CDM parameter space, it can be seen that $P_{\rm ZI}$ gives the tightest constraints on $f_{\rm NL}^{\rm local}$ and $f_{\rm NL}^{\rm equil}$, improving substantially over the bispectrum and over the joint $P_\delta\oplus B_\delta$ constraints. For $f_{\rm NL}^{\rm ortho}$, the scale-dependent peak/valley statistics give the smallest error \citep{Wang2025}, while $P_{\rm ZI}$ remains comparable to $P_\delta\oplus B_\delta$. $P_{\rm ZI}$ also provides competitive constraints on the standard $\Lambda$CDM parameters, with particularly strong performance for $h$, $n_s$, and $\Omega_m$. 

In the $M_\nu+\Lambda$CDM parameter space, $P_{\rm ZI}$ gives the tightest constraint on the summed neutrino mass among the statistics listed in Table~\ref{tab:constraints_mnu}. It also remains highly competitive for the $\Lambda$CDM parameters, yielding the smallest errors for $h$, and matching the best constraints on $n_s$ and $\Omega_b$. The combination of halo mass function and void size function \citep{Bayer2021} gives the tightest constraint on $\Omega_m$, while the wavelet scattering transform \citep{Valogiannis2022} gives the tightest constraint on $\sigma_8$. 

These comparisons should be interpreted as an idealized but consistent benchmark based on real-space matter fields. Observational effects such as galaxy bias, redshift-space distortions, survey geometry, and even baryonic physics will degrade the constraining power of all summary statistics, not only $P_\mathrm{ZI}$. In this benchmark, $P_\mathrm{ZI}$ provides the tightest constraints on most parameters, with particularly significant improvements for $f_{\rm NL}^{\rm local}$, $f_{\rm NL}^{\rm equil}$, and $M_\nu$. None of the other statistics compared here are optimal for more than two parameters.

\begin{table}
\caption{\label{tab:constraints_png}
Comparison of marginalized 1$\sigma$ parameter constraints from different summary statistics, with $M_\nu=0$ fixed. $B_\delta$ denotes the matter bispectrum, $P_\delta\oplus B_\delta$ its combination with the matter power spectrum, and scale-VLYDF$\oplus$-PKHF the combination of scale-dependent valley depth and peak height functions. The $B_\delta$ and $P_\delta\oplus B_\delta$ constraints were provided by the authors of \citet{Floss2024}, while the scale-VLYDF$\oplus$-PKHF constraints are taken from \citet{Wang2025}. All results are based on the real-space matter density field of the \textsc{Quijote} suite at $z=0$ with $k_\mathrm{max}=0.5\ h\mathrm{Mpc}^{-1}$. }
\footnotesize
\begin{ruledtabular}
\renewcommand{\arraystretch}{1.}
\begin{tabular}{lcccc}
 &
$B_\delta$ &
$P_\delta\oplus B_\delta$ &
scale-VLYDF$\oplus$-PKHF &
$P_{\mathrm{ZI}}$ 
\\
\colrule
$f_\text{NL}^\text{local}$ & 104.8  & 52.02   & 48.48     & \textbf{10.75}  \\
$f_\text{NL}^\text{equil}$ & 189.7  & 160.6   & 150.0     & \textbf{51.83} \\
$f_\text{NL}^\text{ortho}$ & 84.81  & 49.24   & \textbf{34.13}     & 45.50 \\
$h$                        & 0.17   & 0.089   & 0.18      & \textbf{0.065} \\
$n_s$                      & 0.16   & 0.064   & 0.032     & \textbf{0.021} \\
$\Omega_m$                 & 0.028  & 0.013   & 0.056     & \textbf{0.010} \\
$\Omega_b$                 & 0.014  & \textbf{0.008}   & 0.020     & \textbf{0.008} \\
$\sigma_8$                 & 0.009  & 0.003   & 0.010     & \textbf{0.002} 
\end{tabular}
\end{ruledtabular}
\end{table}

\FloatBarrier
\begin{table}
\caption{\label{tab:constraints_mnu}
Comparison of marginalized 1$\sigma$ constraints for $\{M_\nu,\ h,\ n_s,\ \Omega_m,\ \Omega_b,\ \sigma_8\}$, with PNG amplitudes fixed to zero. The WST, HMF$\oplus$VSF, $M$, $P_\mathrm{web}^\mathrm{T}$, and $P_\mathrm{web}^\mathrm{NEXUS+}$ denote the wavelet scattering transform \citep{Valogiannis2022}, halo mass function plus void size function \citep{Bayer2021}, marked power spectrum \citep{Massara2021}, T-web environment-dependent power spectra \citep{Bonnaire2022}, and NEXUS+-web environment-dependent power spectra \citep{Sunseri2025}, respectively. All results are taken from the corresponding references and use the real-space matter density field of the \textsc{Quijote} suite at $z=0$ with $k_\mathrm{max}=0.5\ h\mathrm{Mpc}^{-1}$. }
\footnotesize
\begin{ruledtabular}
\renewcommand{\arraystretch}{1.}
\begin{tabular}{lcccccc}
 &
WST &
HMF$\oplus$VSF  &
$M$ &
$P_\mathrm{web}^\mathrm{T}$ &
$P_\mathrm{web}^\mathrm{NEXUS+}$ &
$P_{\mathrm{ZI}}$ 
\\
\colrule
$M_\nu$      & 0.008   & 0.096   & 0.017      & 0.036  & 0.010 & \textbf{0.005} \\
$h$          & 0.104   & 0.230   & 0.098      & 0.078  & 0.069 & \textbf{0.068} \\
$n_s$        & 0.031   & 0.100   & 0.048      & 0.031  & \textbf{0.026} & \textbf{0.026} \\
$\Omega_m$   & 0.014   & \textbf{0.006}   & 0.013      & 0.012  & 0.011 & 0.011 \\
$\Omega_b$   & 0.012   & 0.037   & 0.010      & 0.009  & \textbf{0.008} & \textbf{0.008} \\
$\sigma_8$   & \textbf{0.001}   & 0.007   & 0.002      & 0.002  & 0.002 & 0.002
\end{tabular}
\end{ruledtabular}
\end{table}

\FloatBarrier

\bibliography{paper}{}
\bibliographystyle{aasjournalv7}



\end{CJK*}
\end{document}